\documentstyle[pre,aps,epsf,multicol]{revtex}

\begin{document}

\title{
Unusual beats of the perpendicular-current
 giant magneto-resistance\\and thermopower
 of magnetic trilayers\vspace{1cm}
}

\author{S.~Krompiewski$^1$ and
 U. Krey$^2$ %{\footnote{corresponding author}}
}
\address{$^1$ Institute of Molecular Physics,
 P.A.N., Smoluchowskiego 17, 60-179 Pozna\'n, Poland \\
${}^2$ Institut f\"ur Physik II, Universit\"at Regensburg, 93040
 Regensburg, Germany
         }
\maketitle
\centerline {(received July 1996)}

\begin{abstract}
Oscillations of the giant magnetoresistance (GMR) and thermo-electric
power (TEP) {\it vs.} both the thickness of the non-magnetic spacer
and also that of the ferromagnetic slabs are studied in the
current-perpendicular-to-plane (CPP) geometry of magnetic trilayer
systems, in terms of a single-band tight-binding model without
impurities. The spin-dependent conductance has been calculated from
the Kubo formula by means of a recursion Green's function method and
the semi-infinite ideal-lead wires trick. Additionally the TEP is
obtained directly from  Mott's formula. In general,
the thickness oscillations of the GMR and the TEP may have just one or
two (short and long) oscillations. The long period, related to
spectacular beats, is apparently of non-RKKY type. The TEP
oscillations are strongly enhanced with respect to those of the GMR,
have the same periods, but different phases and a negative bias.
 \end{abstract}
\vspace*{4mm}

\begin{multicols}{2}
%\section{Introduction}

 The phenomenon of giant magnetoresistance ({\it
GMR}) in magnetic multilayers has been intensively studied both
experimentally \cite{l:ba,l:bi} and theoretically
\cite{l:ca,l:le,l:sche,l:as,l:ma} for more than five years now. After
the paper \cite{l:sche} was published it has become clear that one can
expect a large magnetoresistance even in systems which have no
impurities and no structural defects. As pointed out in \cite{l:ma},
the {\it GMR}-oscillations have not only {\it RKKY}- type components,
but may additionally reveal large particular non-{\it RKKY}
oscillations (see below). For the case of semi-infinite ferromagnetic
slabs, a period of those extra oscillations with the {\it spacer
thickness} in the {\it CPP} (current perpendicular-to-plane)
geometry has been shown in \cite{l:ma} to originate from values of the
in-plane wavenumber $\vec{k}_{\|}$ at which at least one spectral
density  vanishes at the ferromagnetic interface.

In the present letter we study the {\it CPP-GMR}
(current-perpendicular-to-plane) behaviour of layered systems of the
type $ W/F_1/S/F_2/W$, where {\it W} stands for a semi-infite ideal
lead wire, $ F_1$ and $ F_2$ for different ferromagnets, generally
with different thicknesses, and {\it S} for the non-magnetic
spacer. Thus, apart from the leads, we are dealing with the usual
magnetic trilayer systems.  Our aim is to show that very pronounced
beats may also, under some circumstances, appear as a function of the
{\it thicknesses of the ferromagnetic slabs}.  Besides the giant
magneto-resistance effect, we also calculate the corresponding effect
for the thermo-electric power ({\it TEP}). 

% \section{Basic Definitions}
  
The calculation technique, we have
developed, is based on the Green's function recursion method
\cite{l:lee,l:as}; it is close to that of \cite{l:as} with one
essential exception, namely instead of working with {\it finite}
systems in the in-plane space (x, y directions), we have reduced the
whole problem to one dimension by performing the Fourier transform in
the {\it infinite} x-y plane. Hence, our Green's functions fulfil the
following equation:

\begin{equation}\label{eqn1}
\sum_{z''} \left[ (E_F-A^{\sigma}_{z})\delta_{zz''} - t_{zz''} \right]
G_{\sigma}(\vec{k}_{\|},z'',z') = \delta_{zz'} \, ,
\end{equation}

where for the simple cubic lattice $t_{z,z'}=t\,\delta_{z',z\pm 1}$ 
and
\begin{equation}\label{eqn2} A_z^{\sigma} = V_{\sigma}(z) +2 t\cdot
[\cos (k_x\cdot a) + \cos (k_y\cdot a)]\, . \end{equation}

In eqn.(\ref{eqn2}), $E_F$ is the Fermi energy; 
$t$ ($<0$) is the nearest-neighbour hopping integral;
$a$ is the lattice constant,
 and $V_\sigma(z)$ is the spin-dependent atomic
potential, which we assume to be constant for given material.
 Hereafter, both $|t|$ and the lattice constant are taken as
energy- and length units, respectively, which formally corresponds to
$t=-1$ and $a$=1. The conductance of electrons of spin $\sigma$ is
given by the well-known Kubo formula, which in turn may be expressed
in terms of Green's functions. In our case the relevant Green
functions are defined by eqn.~(\ref{eqn1}), and the conductance reads:

\begin{eqnarray} 
\Gamma_{\sigma}&=& \frac{8 e^2 N_x 
N_y}
{h\,(2\pi)^2} \int_{BZ} d^2{k}_{\|}\cr
\label{eqn3} & & [ G_{\sigma}''(i,i)
G_{\sigma}''(i-1,i-1) -G_{\sigma}''(i,i-1)^2 ]\,.
\end{eqnarray}

Here the index $i$, which counts the $z$-planes, is arbitrary, since
the same amount of current flows through every cross-section. $G''$
stands for the imaginary part of $G$, and the integration is over the
2-dimensional Brillouin zone. We refer the reader to {\it Refs.}
\cite{l:lee,l:as} for the detailed description how the Green functions
of interest can be found from recursion equations, and we only mention
that due to the ideal-leads trick, the first step of the iteration is
analogous to what is known as the square-root terminator in the
continuous fractions method \cite{l:be}. The prefactor in
Eq.~(\ref{eqn3}) contains the conductance quantum $e^2/h$ as well as
the cross-section area $N_x N_y$ to which the number of active
conductivity channels is proportional.

Due to this effective reduction of the problem to one dimension, we
could easily compute, by a rapid recursion procedure, the Greens
functions in eq.~(\ref{eqn3}) for a large number of
$\vec{k}_{\|}$-vectors and afterwards perform the integration of
eqn.~(\ref{eqn3}) very accurately by means of the highly efficient
{\it special k-points} method of Cunningham \cite{l:cu}. The {\it GMR}
has been defined as

\begin{equation} \label{eqn4}
GMR = \frac{ \Gamma^{\uparrow\uparrow}_{\uparrow} +
 \Gamma^{\uparrow\uparrow}_{\downarrow}}{
 \Gamma^{\uparrow\downarrow}_{\uparrow} + 
\Gamma^{\uparrow\downarrow}_{\downarrow}}
-1 \, ,
\end{equation}

where the {\it superscripts} of $\Gamma$ refer to the parallel and
antiparallel magnetization configurations of the two ferromagnets,
whereas the {\it subscripts} refer to the carriers considered.

%\section{Results}
%\subsection{Systems considered}
 Before we performed original calculations, we have tested our
programs by considering systems studied by Asano {\it et al.} in
\cite{l:as}, namely: (i) a non-magnetic multilayer {\it /A/B/A/B}
consisting of 4 sublayers, each three monolayers thick with
alternating $\pm 1$ potentials, and (ii) their model for the {\it
FeCr} superlattice: We have found very good agreement concerning both
the {\it CPP-GMR} per unit area as well as the densities of states at
the Fermi level, although the system of \cite{l:as} has got only $12
\times 12$ atoms in the x-y plane and uses free boundary conditions.

Our results for the {\it CPP-GMR} of the $W/F_1/S/F_2/W$ systems are
presented in the figures below. The plots correspond to the atomic
potentials equal zero for both up- and down-spin electrons in the
spacer and in the infinite lead wires. In the ferromagnets, the
potentials for electrons with spin $\sigma=\pm 1$ take the following
values:
$V_{1\sigma}$ and $V_{2\sigma}$ for the $\uparrow\uparrow$-configuration 
and $V_{1\sigma}$, and $V_{2(-\sigma})$ for the $\uparrow
\downarrow$-configuration, 
 where $V_{1\sigma}$ resp.~$V_{2\sigma}$ and
$V_{2(-\sigma)}$ concern the slabs $ F_1$ and $ F_2$,
respectively. (The actual values are given below and are also denoted
in the figure captions). Additionally we have assumed a perfect
matching of the minority bands of the ferromagnets with the spacer
bands by putting $V_{1\downarrow}=V_{2\downarrow}=0$ ({\it
cf.}~\cite{l:as}). Finally, unless otherwise stated, we have chosen
the Fermi energy $E_F$=2.5 well above the potential barriers, see
below.

 It can be seen from Fig.1 and Fig.2 
 that the {\it CPP-GMR} oscillates as a function of the thickness
$n_f$ of {\it one or both ferromagnetic sandwiches}
 with a short period of $\approx 2$ monolayers (ML).
Additionally, in Fig.1, but not in Fig.2, {\it pronounced beats} are
observed with typical repetition lengths of roughly $10$ ML. Apparently
those beats always appear, if the depth of the potential well
corresponding to the ferromagnetic sandwich(es) with varying thickness
is $V_{2\uparrow}=-1.8$ (Fig.1) or $-1$ (not shown), whereas in case
of $V_{2\uparrow}\equiv -2$ the beats are absent (Fig.2).

 In case of Fig.3,
where again $V_{2\uparrow}=-2$, but where only the spacer thickness
$n_s$ is varied, one observes an intermediate period of $\sim 4.5$
ML, whereas  the beats do not appear.

\input epsf
{\epsfxsize=9cm
%\epsfbox{../krs04114/plot/gmrfig1.eps} 
\epsfbox{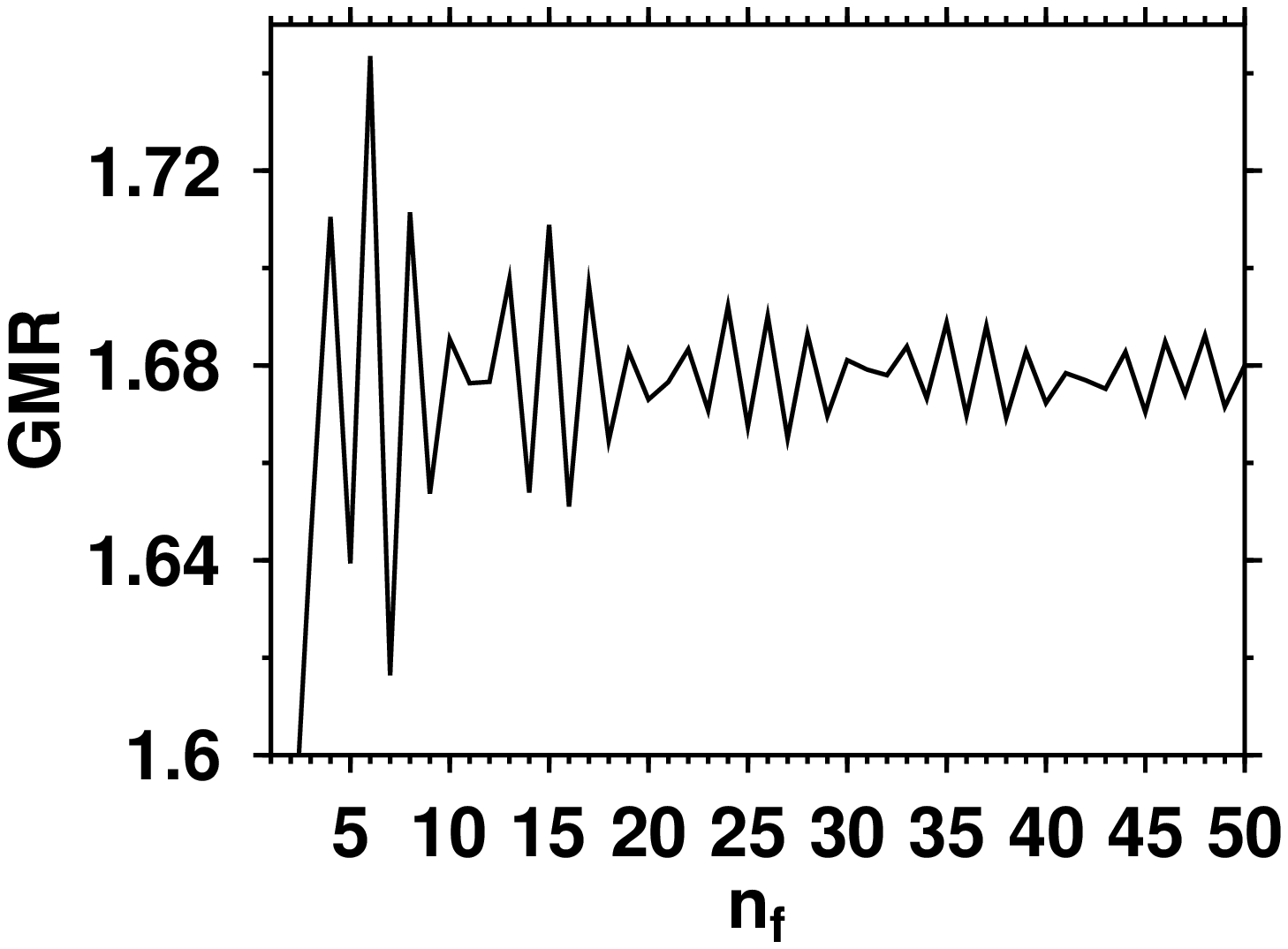}

{\small{FIG.1}: {\it CPP-GMR} of the system $ 3F_1/6S/n_fF_2$
between two infinite lead wires. $V_{1\uparrow}=-2$ and
$V_{2\uparrow}=-1.8$ in the ferromagnets (all other potentials are 0).
$ E_F$=2.5.
 Similar results apply for the system
 $n_fF_1/6S/n_fF_1$ with
$V_{1\uparrow}=V_{2\uparrow}=-1.8$.
}}
\par\vskip 0.2 truecm
{ \epsfxsize=9cm
%\epsfbox{../krs04114/plot/gmrfig2.eps}
\epsfbox{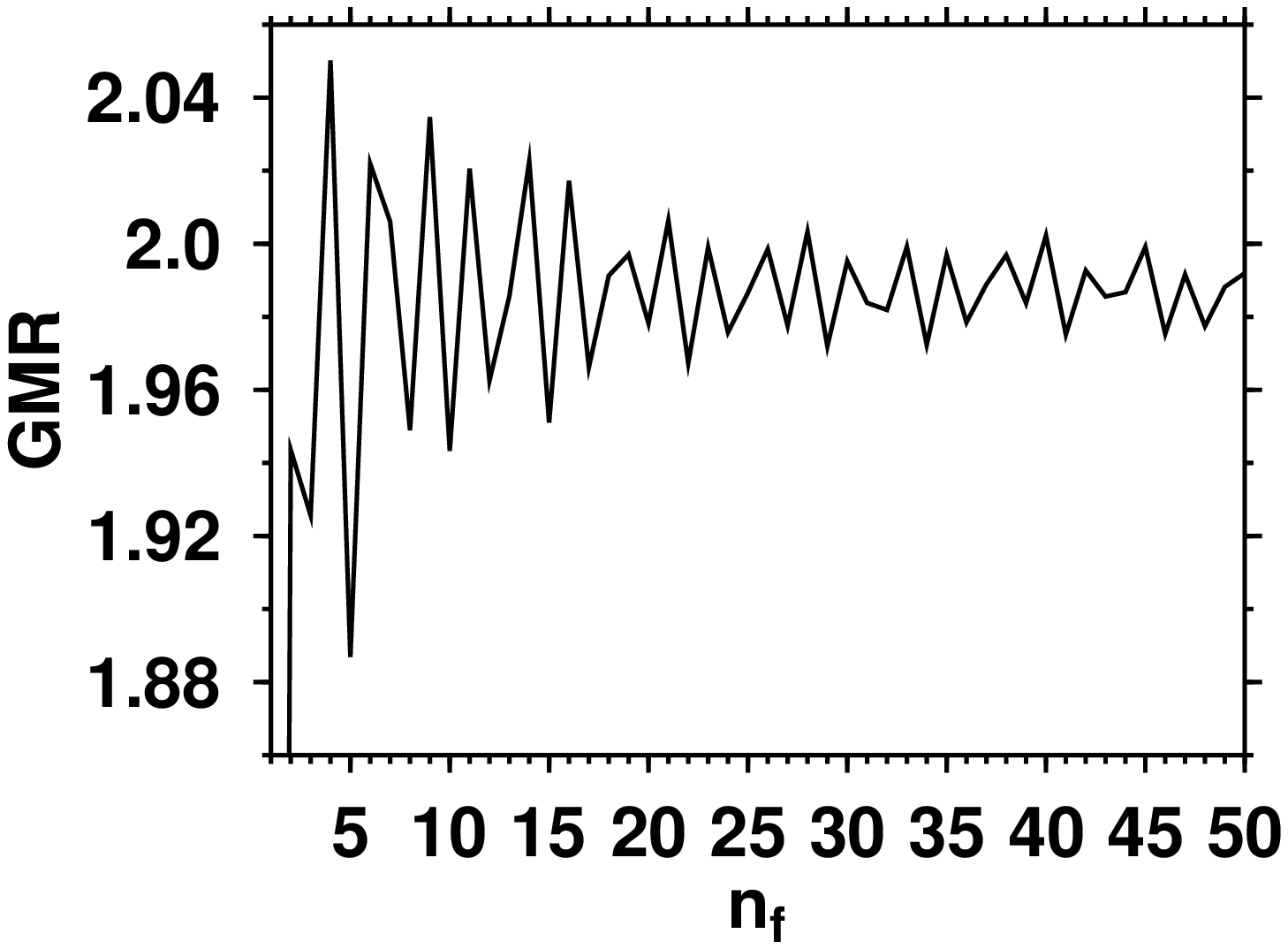} 
{\small{FIG.2} 
   The same as Fig.1, but with
$V_{1\uparrow}=V_{2\uparrow}=-2$. Again similar results apply
 for the system
 $n_fF_1/6S/n_fF_1$ with
$V_{1\uparrow}=V_{2\uparrow}=-2$.}} 

{
\epsfxsize=9cm
%\epsfbox{./stefan/gmrfig3-ns-3.eps}
\epsfbox{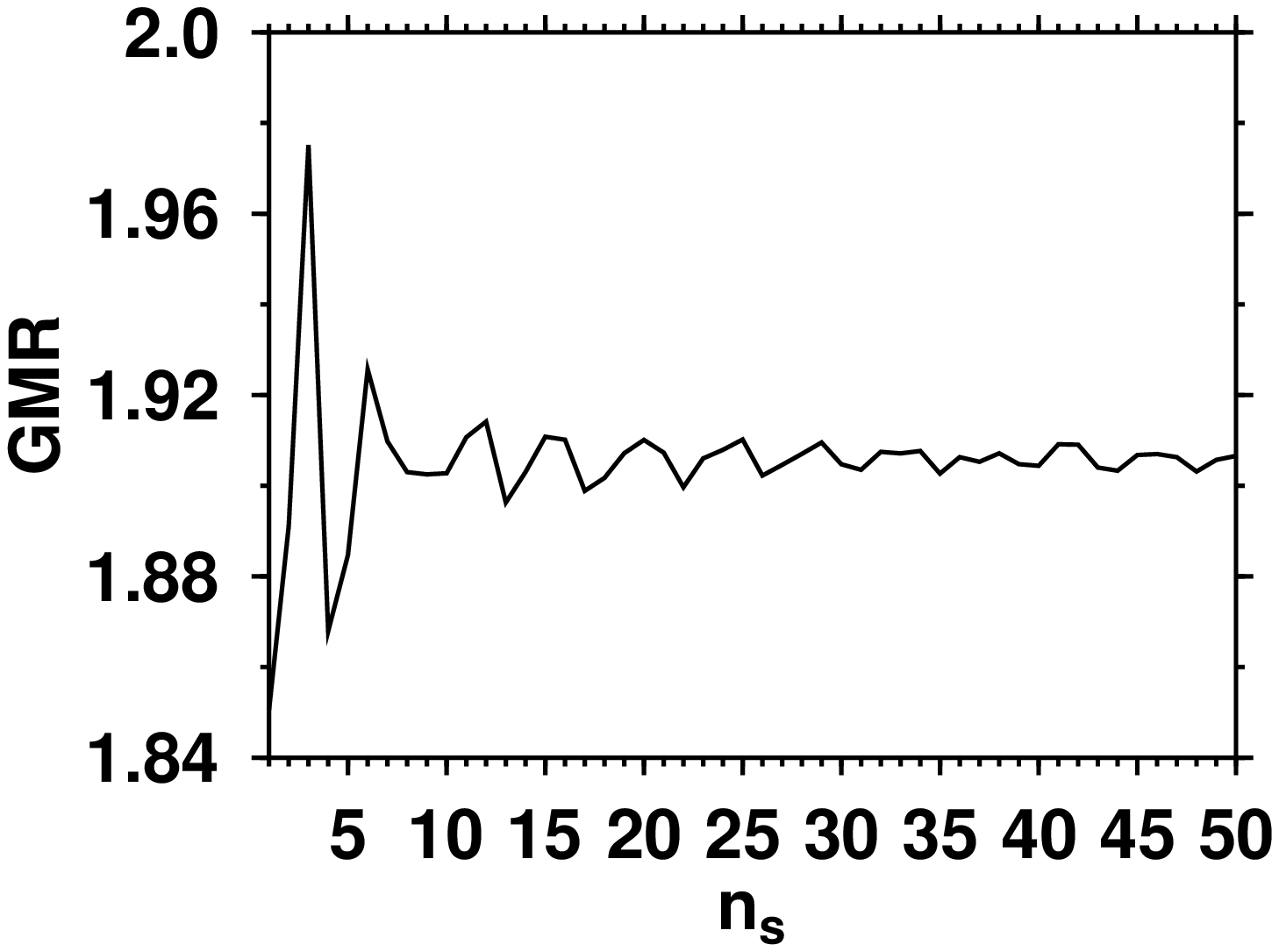}
{\small{FIG.3}: 
The same as Fig.2, but with $n_f$=3 on both sides, as a function
of the spacer thickness $n_s$.
}}

 In any case it can be easily seen that the {\it GMR} oscillations
decay always roughly $\sim n_f^{-1}$, analogously to the findings of
Mathon {\it et al.} concerning the oscillations {\it vs.}~the {\it
spacer thickness} observed in \cite{l:ma}. It results from these
observations that oscillations and beats of the {\it GMR}, similarly
as those of the {\it exchange coupling}, see \cite{l:Euro}, are not
specific for the non-magnetic spacer, but concern the ferromagnetic
[Bslabs, too.

%\subsection{Characteristic wavenumbers}

  In the following we try to identify the {\it RKKY}- and {\it
  non-RKKY} periods visible in our figures, using
arguments along the lines of \cite{l:ma,l:al}.

 In \cite{l:ma} Mathon {\it et al.}~observed roughly analogous
behaviour of the {\it GMR} as a function of the {\it spacer} thickness
in a simpler system, namely a spacer imbedded between two identical
semi-infinite ferromagnets. As in our case, the authors of \cite{l:ma}
assume that within the spacer the potential $V(z)$ vanishes, but
concerning the ferromagnets their assumptions are different, since
they assume the same vanishing potential for electrons with majority
spin, but a {\it positive} potential $V$ for electrons with {\it
minority} spin, so that in their case the spacer represents a
potential well for minority carriers. (In contrast, in our case we
have a {\it negative} potential for {\it majority}-spin carriers in
the ferromagnets, but vanishing potential for minority carriers, so
that in our case the ferromagnets represent quantum wells for majority
carriers, which are interrupted by a barrier represented by our spacer
and bounded from outside by the ideal leads, which have the same
potential as the spacer).  It was shown in \cite{l:ma} that (i) a
continuum of wavenumbers $k_z$ contributes to the conductivity
$\Gamma$, i.e.~the dependence on the varying thickness $L$ was
$\Gamma\sim \int {\rm dk_x dk_y}\,\, c(k_x,k_y)\,\cdot\,\exp
[\,2\,{\rm{i}}\,k_z(k_x,k_y,E_f)\,L] $, and (ii) that dominating
contributions can arise a) from wavenumbers, where $k_z(k_x,k_y;E_f)$
is stationary as a function of ($k_x,k_y$), and b) from particular
cut-off wavenumbers ($k_x,k_y$), where $c(k_x,k_y)$ vanishes
abruptly. These cut-off wavenumbers correspond in our case to unbound
majority carriers which in z-direction have just the minimal kinetic
energy in the wires to evade to infinity.
 The contributions under a) and b) are the ''{\it RKKY}'' and
''{\it non-RKKY}-contributions'' mentioned already above.

At this place one should note in any case that the ''non-RKKY
wavenumbers'' under b) do not play a role in the thickness-dependence
of the {\it exchange coupling} of the ferromagnets across the spacer,
see \cite{l:al}.

 In both cases a) and b), one can estimate the important wave numbers
$k_z$ from the asymptotic equation (valid actually for large
thicknesses $n_f$ and constant potential) 
\begin{equation}\label{eqkz}
k_z(k_x,k_y;E_f)=\arccos [\frac{V_\sigma -E_f}{2}-\cos k_x -\cos k_y]\,.
\end{equation}
  Here, in case a), with
$E_F$=2.5 and the values ($k_x, k_y$)=($\pm \pi, \pm \pi$), where
$\cos k_x +\cos k_y$ is extremal, we get $k_z^{(1)}=0.722$ for the
carriers with $V_\sigma =0$ (i.e. for the down-spin and spacer
carriers in case of aligned ferromagnets), and $k_z^{(2)}=1.721$ for
the up-spin carriers with $V_{\uparrow}=-1.8$ and $k_z^{(2)}=1.823$
for $V_\uparrow=-2$.  The corresponding wavelengths can then be
calculated as in \cite{l:al} from the expression $\lambda(p,q) =
\pi/(p k_z^{(1)}+q k_z^{(2)})$, which yields $\lambda(0,1)=1.825$
($V=-1.8$) and =1.723 ($V=-2$), i.e.~$\sim$ the short period of two
monolayers (2 ML) seen in the figures. It is interesting that in
Fig.3, where only the spacer thickness is varied, only a longer
wavelength of $\sim 4.5$ ML is visible, which would correspond to
$\lambda(1,0)= 4.35$. However, these particular ''RKKY-wavenumbers''
would apparently not produce the beats in Fig.1, which
have a much larger repetition length of $\approx 10 ML$.

In order to get the beats seen in Fig.\ 1 and in Fig.4\ below, one
needs apart from the wave characterized by $k_z^{(2)}$ another wave with
a wavenumber very  close to it. As there seems no  way to get such a
wave from eqn.\ (5) with $V_\uparrow=-1.8$, we suggest it to be of {\it
non}-RKKY character.
 This means, it should come from electrons at
 $E=E_f$ with values of $(k_x,k_y)\ne (\pm\pi ,\pm\pi)$. 
 More insight into the nature of {\it non}-RKKY oscillations can
 perhaps only be gained by taking into account localized and resonant
 states in addition to the extended ones used for the sketchy
 estimations of the RKKY wavenumbers above: In fact, although the
 localized states do not contribute directly to the conductance, they
 modify the density of states at $E_f$ and thereby influence the
 conductivity.

%\subsection{Thermo-electric power}

In addition to the conductance $\Gamma(E_f)$, where $E_f$ is the Fermi
energy, we also calculated the thermopower $S$:

\epsfxsize=9cm \epsfbox{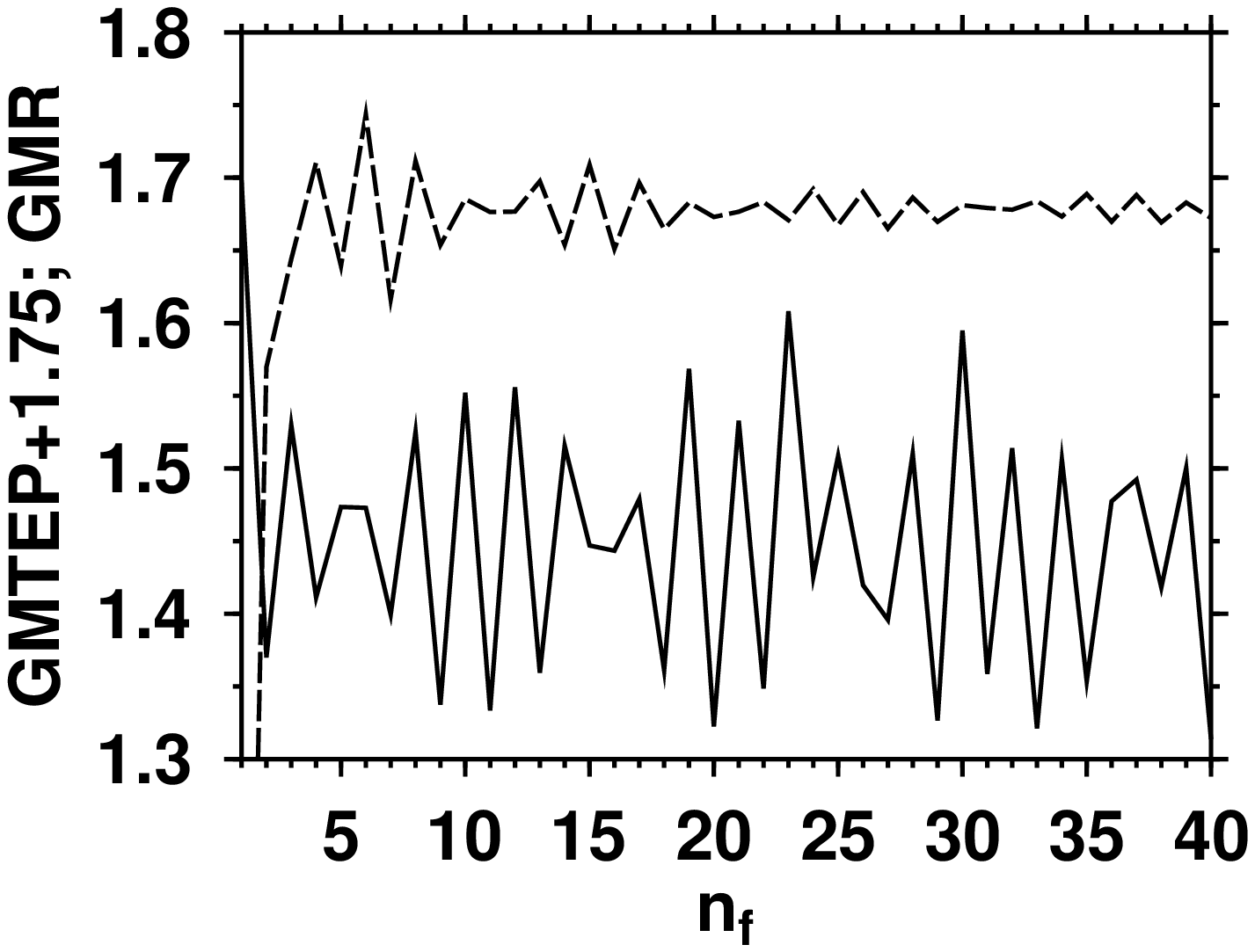}
 {\small{FIG.4}:
 The same as Fig.1, with additional results for
the ''{\it GMTEP}'' (Giant Magneto-{\it TEP} effect, lower curve),
where {\it TEP} means ''thermo-electric power''. Note the shift of the
{\it TEP} results by +1.75.
}

$S(E_f)$ is related to $\Gamma (E_f)$ by  Mott's  formula,
 \cite{l:mott}:
 \begin{equation} \label{eqn6}
S_\sigma = -\frac{\pi^2 k_B^2 T }{3\, |e|}
{\rm (d/dE)} \log \Gamma_\sigma (E)\end{equation}
    at $E=E_f$, where $\sigma$ denotes the spin of the carriers, $T$
the Kelvin temperature, $k_B$ Boltzmann's constant and $e$ the
elementary charge.

Analogously to the GMR, see eqn.~(\ref{eqn4}), we define the
 ''Giant Magneto-{\it TEP}-effect'' {\it GMTEP}, where
{\it TEP} means thermo-electric power, by

\begin{equation} \label{eqn7}
 GMTEP= \frac{ S^{\uparrow\uparrow}_{\uparrow} +
 S^{\uparrow\uparrow}_{\downarrow}}{
 S^{\uparrow\downarrow}_{\uparrow} + 
 S^{\uparrow\downarrow}_{\downarrow}}
-1\,\,.
\end{equation}

It can be seen in Fig.4 that the oscillations of the GMTEP are even
stronger than those of the {\it GMR}, although they have a significant
and negative bias and a different phase than those of the GMR.

%\subsection{Roughness sensitivity}

 Finally it should be noted that the amplitude of the oscillations
observed in the figures 1 and 2 is typically of the order of 1\% to 2\%
only: Thus the oscillations can only be detected by very accurate
calculations as the present one, and only with interfaces of suffcient
quality: In fact, we have ''smeared'' the results for GMR(i) as
follows 
\begin{eqnarray}\label{eqn5} 
 GMR(i) \to (x/2) &\cdot& [GMR(i+1)+GMR(i-1)]\cr 
+(1-x) &\cdot& GMR(i)\,\,,
\end{eqnarray} 
with $x=0.5$ and $x=0.25$, respectively, and obtained
instead of Fig.1 the ''smeared'' Fig.5, where the oscillations
are no longer visible for for strong interdiffusion ($x=0.5$), whereas
for $x=0.25$, they are roughly reduced by a factor (1/2) and can 
still be recognized.

{\epsfxsize=9cm 
%epsfbox{./stefan/gmrfig1b.eps}
\epsfbox{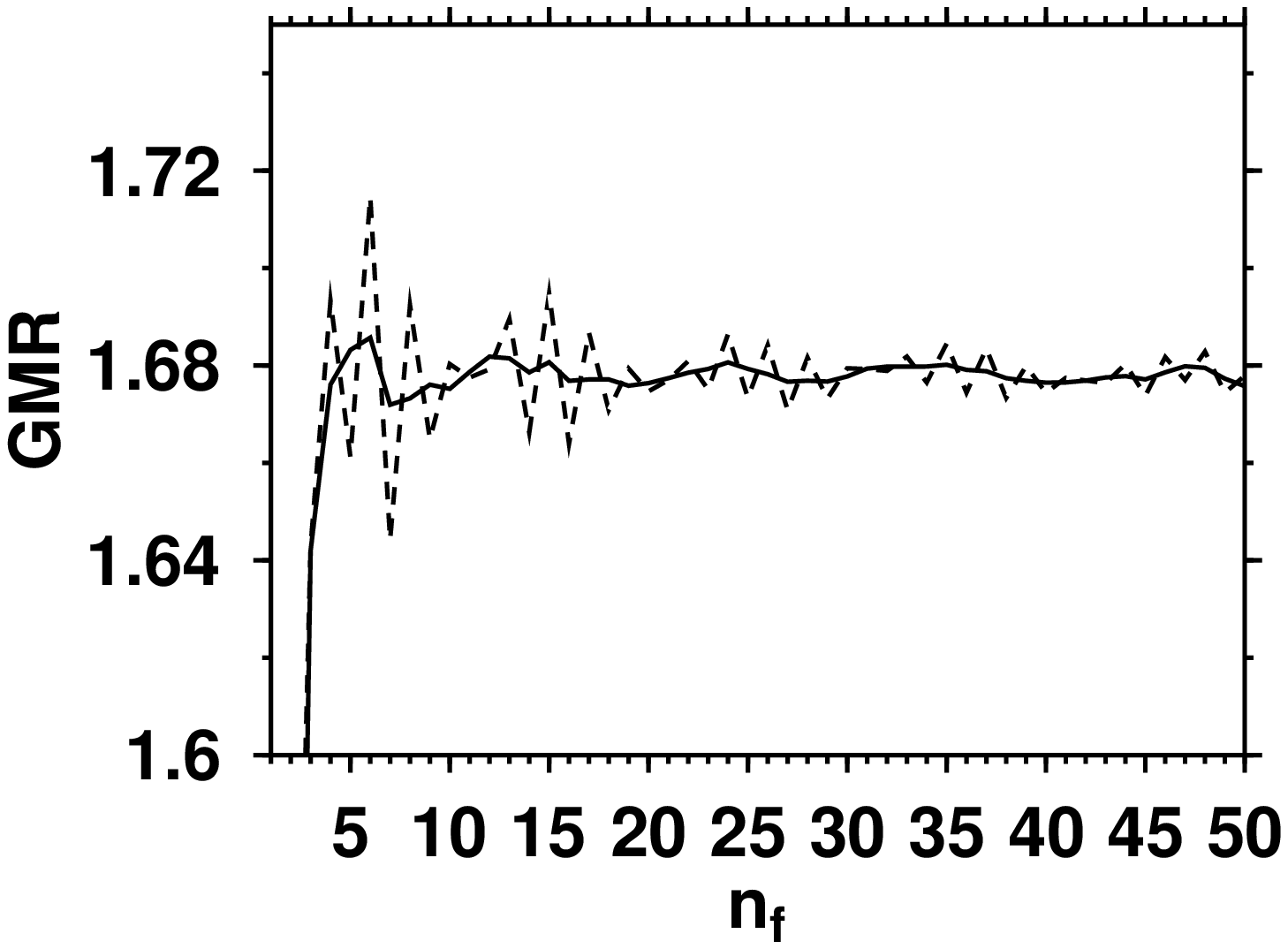}
{\small{FIG.5}: 
The same
as Fig.1, but smeared in accordance with eqn.~(\ref{eqn5})
with $x=0.25$ (dashed line) and $x=0.5$ (solid line).
}}

%\section{Conclusions} 

In conclusion, within an s-band tight-binding
model combined with the Green's function recursion method, we have
reduced the calculation of the {\it CPP-GMR} effectively to one
dimension. We have computed the {\it CPP-GMR} for pure systems
composed of two different ferromagnets separated by a non-magnetic
spacer and sandwiched between ideal infinite lead wires. Unusual beats
of the {\it GMR vs.}~the thickness of the ferromagnetic slabs have
been observed and shown to depend strongly on the kind of
ferromagnetic slabs (i.e.~on the potentials $V_{\sigma}$ representing
the exchange splittings). Similar results, with even stronger beats,
have also been obtained for the ''Giant Magneto-{\it TEP}-Effect'',
where ''{\it TEP}'' means the Thermo-Electric Power. The beats of the
{\it GMR} have been shown to remain visible, if there is only moderate
chemical interdiffusion in the boundary layers. Finally, it has been
shown that the short periods of 2 monolayers visible in the results
arise from the usual ''RKKY''-type mechanism, whereas the beats arise
from certain ''non-RKKY'' cut-off
 contributions, analogous to those suggested in \cite{l:ma}.

%\vspace{0.5cm}

{\bf Acknowledgements}

This work has been carried out under 
the bilateral project DFG/PAN 436 POL and the KBN grants No.~2P 302
 005 07 and 2P 03B 165 10 (S.K.). We also thank the  Pozna\'n,
Munich and Regensburg Computer Centres for computing time.

\end{multicols}
\end{document}